\documentclass[aps,prd,twocolumn,superscriptaddress,showpacs]{revtex4}
\usepackage{epsfig,epsf}
\usepackage{amsmath}
\usepackage{amsthm}
\usepackage{amsfonts}
\usepackage{amssymb}
\usepackage{dsfont}
\usepackage{multirow}
\usepackage{appendix}
\usepackage{slashed}
\usepackage[active]{srcltx}
\usepackage{psfrag}

\begin{document}

\title{ Charmed partner of the exotic $X(5568)$ state and its properties}
\date{\today}
\author{S.~S.~Agaev}
\affiliation{Department of Physics, Kocaeli University, 41380 Izmit, Turkey}
\affiliation{Institute for Physical Problems, Baku State University, Az--1148 Baku,
Azerbaijan}
\author{K.~Azizi}
\affiliation{Department of Physics, Do\v{g}u\c{s} University, Acibadem-Kadik\"{o}y, 34722
Istanbul, Turkey}
\author{H.~Sundu}
\affiliation{Department of Physics, Kocaeli University, 41380 Izmit, Turkey}

\begin{abstract}
The mass, decay constant and width of a hypothetical charmed partner $X_c$
of the newly observed exotic $X_b(5568)$ state are calculated using a
technique of QCD sum rule method. The $X_c=[su][\overline{c}\overline{d}]$
state with $J^{P}=0^{+}$ is described employing two types of the diquark-antidiquark
interpolating currents. The evaluation of the mass $m_{X_c}$ and
decay constant $f_{X_c}$ is carried out utilizing the two-point sum rule method by
including vacuum condensates up to eight dimensions. The widths of the decay
channels  $X_c \to D_s^{-} \pi^{+}$ and $X_c \to D^{0} K^{0}$ are also found. To
this end, the strong couplings $g_{X_{c}D_{s}\pi}$ and $g_{X_{c}D K}$ are computed
by means of QCD sum rules on the light-cone and soft-meson approximation.
\end{abstract}

\maketitle


\section{Introduction}

\label{Sec:Int}

The D0 Collaboration recently reported the observation of a narrow structure
$X_{b}(5568)$ in the decay process $X_{b}(5568) \to B_{s}^{0} \pi^{\pm}$, $%
B_{s}^{0} \to J/\psi \phi$, $J/\psi \to \mu^{+} \mu^{-}$, $\phi \to
K^{+}K^{-}$ based on $p\bar{p}$ collision data at $\sqrt{s}=1.96\ \mathrm{TeV%
}$ collected at the Fermilab Tevatron collider \cite{D0:2016mwd}. The new
state $X_b(5568)$ is considered to posses quantum numbers $J^{PC}=0^{++}$.
The D0 Collaboration provides the value $m_{X_{b}}=5567.8 \pm 2.9 \mathrm{%
(stat)}^{+0.9}_{-1.9} \mathrm{(syst)}\, \mathrm{MeV}$ for its mass, and
estimates $\Gamma=21.9 \pm 6.4 \mathrm{(stat)}^{+5.0}_{-2.5} \mathrm{(syst)}%
\, \mathrm{MeV}$ for its decay width. As it was emphasized in Ref.\ \cite%
{D0:2016mwd} this is the first observation of a hadronic state with four
different quark flavors. Thus, $X_b$ state is composed of $b,\, s, \, u, \, d
$ quarks.

Suggestions concerning the possible quark-antiquark organization of $X_b$
were made already in Ref.\ \cite{D0:2016mwd}. Thus, within the
diquark-antidiquark model, the $X_b$ state with positive charge, i.e. the
particle $X_b^{+}$ may be described as $[bu][\bar{d}\bar{s}]$ or $[su][\bar{b%
}\bar{d}]$ bound states, whereas $X_b^{-}$ may have the structures $[bd][%
\bar{s}\bar{u}]$ or $[sd][\bar{b}\bar{u}]$. Alternatively, $X_b$ may be
considered as a molecule composed of $B$ and $\overline{K}$ mesons.

This is a valuable discovery, because the charmonium-like resonances that
populate $XYZ$ family of "traditional" exotic states, contain $c\bar{c}$
charm quark-antiquark pair and hence, the number of the quark flavors in
these particles does not exceed three. Properties of known exotic states
extracted from experimental data and theoretical calculations can be found
in, for instance, review papers \cite%
{Swanson:2006st,Klempt:2007cp,Godfrey:2008nc,Voloshin:2007dx,Nielsen:2010,
Faccini:2012pj,Esposito:2014rxa,Chen:2016} and references therein.

The newly observed state $X_b(5568)$ has immediately attracted interests of
physicists and stimulated theoretical studies of $X_b$ in the context of
different approaches \cite{Agaev:2016mjb,Agaev:2016ijz,Wang:2016mee,Wang:2016tsi,Chen:2016mqt,Zanetti:2016wjn,Xiao:2016mho,Liu:2016ogz,Liu:2016xly}.
 Thus, in Refs.\ \cite{Agaev:2016mjb,Agaev:2016ijz} we
have calculated the mass, decay constant and width of the $X_b(5568)$ state
within the diquark-antidiquark picture $X_b=[su][\bar{b}\bar{d}]$
considering the exotic state with positive charge. Our predictions for the
mass $m_{X_b}$, and for the width of its decay $\Gamma(X_{b}^{+}\to
B_{s}^{0}\pi^{+})$ are in agreement with the experimental data. It is worth
noting that in the context of the diquark model some parameters of $X_b$
were also analyzed in Refs.\ \cite%
{Wang:2016tsi,Chen:2016mqt,Zanetti:2016wjn,Wang:2016mee}. In these works the
authors use various versions for the diquark-antidiquark type interpolating
currents with different Lorentz structures. It is remarkable, that the
obtained values for $m_{X_{b}}$ are in agreement with each other and also
consistent with experimental data of D0 Collaboration. The molecule picture
for $X_b$ was realized in Ref.\ \cite{Xiao:2016mho}, where the $X_b(5568)$
state was taken as the $B\overline{K}$ bound state. The questions of
quark-antiquark organization of this particle and its partners were
addressed in Ref.\ \cite{Liu:2016ogz}.

In the present work we are going to continue our investigation of the new
family of the four-quark exotic states by considering the hypothetical
charmed partner of the $X_b(5568)$ state, which is composed of the $c,\,s,\,
u,\, d$ quarks. We assume that this state bears the same quantum numbers as
its counterpart, i.e. $J^{PC}=0^{++}$. We also accept that it has the
internal structure $X_c=[su][\bar{c}\bar{d}]$ in the diquark-antidiquark
model. Thus, the partner state $X_c$ is a neutral particle. Our aim is to
determine the parameters of the state $X_c$, i.e. to find its mass, decay
constant and widths of the strong $X_{c}\to D_{s}^{-}\pi^{+}$ and
$X_{c}\to D^{0}K^{0}$ decays. For these purposes, we apply methods presented in a rather
detailed form in Refs.\ \cite{Agaev:2016mjb,Agaev:2016ijz,Agaev:2016dev}.

This work is structured in the following way. In Section \ref{sec:Mass} we
introduce the interpolating currents employed in QCD sum rule calculations.
Here we find the mass and decay constant of $X_c$ using
the two-point QCD sum rule approach. The widths of the strong decays $X_{c}\to D_{s}^{-}\pi^{+}$
and $X_{c}\to D^{0}K^{0}$ are subject of Sect.\ \ref{sec:Vertex}.
Explicit expression for the spectral density required in computation of the
mass and decay constant of the exotic $X_c$ is moved to Appendix A.


\section{The mass and decay constant of $X_c$}

\label{sec:Mass}

As it has been noted above, we use the two-point QCD sum rule approach in order to
compute mass and decay constant of the $X_{c}$ state. To this end, we
consider the two-point correlation function given as
\begin{equation}
\Pi (p)=i\int d^{4}xe^{ip x}\langle 0|\mathcal{T}%
\{J_{1(2)}^{X_{c}}(x)J_{1(2)}^{X_{c}\dag }(0)\}|0\rangle ,  \label{eq:CorrF1}
\end{equation}%
where $J_{1(2)}^{X_{c}}(x)$ are the interpolating currents with required
quantum numbers. We consider $X_{c}$ state as a particle with the quantum
numbers $J^{PC}=0^{++}$. Then in the diquark-antidiquark model the current $%
J_{1}^{X_{c}}(x)$ is given by the following expression
\begin{equation}
J_{1}^{X_{c}}(x)=\varepsilon ^{ijk}\varepsilon ^{imn}\left[ s^{j}(x)C\gamma
_{\mu }u^{k}(x)\right] \left[ \overline{c}^{m}(x)\gamma ^{\mu }C\overline{d}%
^{n}(x)\right].  \label{eq:CDiq1}
\end{equation}
Alternatively, one may introduce the interpolating current
\begin{equation}
J_{2}^{X_{c}}(x)=\varepsilon ^{ijk}\varepsilon ^{imn}\left[ s^{j}(x)C\gamma
_{5 }u^{k}(x)\right] \left[ \overline{c}^{m}(x)\gamma_{5}C\overline{d}^{n}(x)%
\right].  \label{eq:CDi2}
\end{equation}
In Eqs.\ (\ref{eq:CDiq1}) and (\ref{eq:CDi2}) $i,\ j,\ k,m,\ n$ are color
indexes and $C$ is the charge conjugation matrix.

Let us note that the current $J_{1}^{X_{c}}(x)$ has been employed throughout
in Refs.\ \cite{Agaev:2016mjb,Agaev:2016ijz} for exploration the exotic $%
X_b(5568)$ state. The sum rules derived in these works, after trivial
replacements of corresponding parameters, can easily be applied to analyze
the $X_c$ state. Therefore, in what follows we concentrate on the current $%
J_{2}^{X_{c}}(x)$ omitting, in what follows, the subscript in its definition.

The representation of the function $\Pi (p)$ in terms of the physical
quantities does not depend on the form of the interpolating current and is
the same for both $J_{1(2)}^{X_{c}}(x)$
\begin{equation}
\Pi ^{\mathrm{Phys}}(p)=\frac{\langle 0|J^{X_{c}}|X_{c}(p)\rangle \langle
X_{c}(p)|J^{X_{c}\dag }|0\rangle }{m_{X_{c}}^{2}-p^{2}}+...
\label{eq:CorrF1A}
\end{equation}%
where $m_{X_{c}}$ is the mass of the $X_{c}$ state, and dots stand for
contributions of the higher resonances and continuum states. We define the
decay constant $f_{X_{c}}$ through the matrix element%
\begin{equation}
\langle 0|J^{X_{c}}|X_{c}(p)\rangle =f_{X_{c}}m_{X_{c}}.
\end{equation}
Then, for the correlation function we obtain
\begin{equation}
\Pi ^{\mathrm{Phys}}(p)=\frac{m_{X_{c}}^{2}f_{X_{c}}^{2}}{m_{X_{c}}^{2}-p^{2}%
}+\ldots  \label{eq:CorM}
\end{equation}%
The Borel transformation applied to Eq.\ (\ref{eq:CorM}) yields
\begin{equation}
\mathcal{B}_{p^{2}}\Pi ^{\mathrm{Phys}%
}(p)=m_{X_{c}}^{2}f_{X_{c}}^{2}e^{-m_{X_{c}}^{2}/M^{2}}+\ldots
\label{eq:CorBor}
\end{equation}

The theoretical expression for the same function, $\Pi ^{\mathrm{QCD}}(p)$,
has to be determined employing the quark-gluon degrees of freedom.
Contracting the quark fields we find for the correlation function $\Pi ^{%
\mathrm{QCD}}(p)$ :
\begin{eqnarray}
&&\Pi ^{\mathrm{QCD}}(p)=i\int d^{4}xe^{ipx}\varepsilon ^{ijk}\varepsilon
^{imn}\varepsilon ^{i^{\prime }j^{\prime }k^{\prime }}\varepsilon
^{i^{\prime }m^{\prime }n^{\prime }}  \notag \\
&&\times \mathrm{Tr}\left[ \gamma _{5}\widetilde{S}_{d}^{n^{\prime
}n}(-x)\gamma _{5}S_{c}^{m^{\prime }m}(-x)\right]  \notag \\
&&\times\mathrm{Tr}\left[ \gamma _{5}\widetilde{S}_{s}^{jj^{\prime
}}(x)\gamma_{5 }S_{u}^{kk^{\prime }}(x)\right] .  \label{eq:CorrF2}
\end{eqnarray}
where $S_{q}^{ij}(x)$ and $S_{c}^{ij}(x)$ are the light ($q\equiv u,\ d\ $or
$s$) and $c$-quark propagators, respectively. In Eq.\ (\ref{eq:CorrF2}) we
introduce the notation
\begin{equation*}
\widetilde{S}_{q}^{ij}(x)=CS_{q}^{ijT}(x)C.
\end{equation*}%
In the $x$-space the light quark propagator $S_{q}^{ij}(x)$ has the form
\begin{eqnarray}
&&S_{q}^{ij}(x)=i\delta _{ij}\frac{\slashed x}{2\pi ^{2}x^{4}}-\delta _{ij}%
\frac{m_{q}}{4\pi ^{2}x^{2}}-\delta _{ij}\frac{\langle \overline{q}q\rangle
}{12}  \notag \\
&&+i\delta _{ij}\frac{\slashed xm_{q}\langle \overline{q}q\rangle }{48}%
-\delta _{ij}\frac{x^{2}}{192}\langle \overline{q}g\sigma Gq\rangle +i\delta
_{ij}\frac{x^{2}\slashed xm_{q}}{1152}\langle \overline{q}g\sigma Gq\rangle
\notag \\
&&-i\frac{gG_{ij}^{\alpha \beta }}{32\pi ^{2}x^{2}}\left[ \slashed x{\sigma
_{\alpha \beta }+\sigma _{\alpha \beta }}\slashed x\right] -i\delta _{ij}%
\frac{x^{2}\slashed xg^{2}\langle \overline{q}q\rangle ^{2}}{7776}  \notag \\
&& -\delta _{ij}\frac{x^{4}\langle \overline{q}q\rangle \langle
g^{2}GG\rangle }{27648}+ \ldots  \label{eq:qprop}
\end{eqnarray}%
For the $c$-quark propagator $S_{c}^{ij}(x)$ we employ the expression from
Ref.\ \cite{Reinders:1984sr}
\begin{eqnarray}
&&S_{c}^{ij}(x)=i\int \frac{d^{4}k}{(2\pi )^{4}}e^{-ikx}\left[ \frac{\delta
_{ij}\left( {\slashed k}+m_{c}\right) }{k^{2}-m_{c}^{2}}\right.  \notag \\
&&-\frac{gG_{ij}^{\alpha \beta }}{4}\frac{\sigma _{\alpha \beta }\left( {%
\slashed k}+m_{c}\right) +\left( {\slashed k}+m_{c}\right) \sigma _{\alpha
\beta }}{(k^{2}-m_{c}^{2})^{2}}  \notag \\
&&\left. +\frac{g^{2}}{12}G_{\alpha \beta }^{a}G^{a\alpha \beta }\delta
_{ij}m_{c}\frac{k^{2}+m_{c}{\slashed k}}{(k^{2}-m_{c}^{2})^{4}}+\ldots %
\right] .  \label{eq:Qprop}
\end{eqnarray}%
In Eqs.\ (\ref{eq:qprop}) and (\ref{eq:Qprop})
\begin{equation*}
G_{ij}^{\alpha \beta }\equiv G_{a}^{\alpha \beta
}t_{ij}^{a},\,\,\,\,a=1,\,2\,\ldots 8,
\end{equation*}%
where $i,\,j$ are color indexes, and $t^{a}=\lambda ^{a}/2$ with $\lambda
^{a}$ being the standard Gell-Mann matrices. The first term in Eq.\ (\ref%
{eq:Qprop}) is the perturbative propagator of a massive quark, the next two terms are
nonperturbative gluon corrections. In the nonperturbative terms the gluon
field strength tensor $G_{\alpha \beta }^{a}\equiv G_{\alpha \beta }^{a}(0)$
is fixed at $x=0$.

The correlation function $\Pi ^{\mathrm{QCD}}(p^{2})$ is given by a simple
dispersion integral
\begin{equation}
\Pi ^{\mathrm{QCD}}(p^{2})=\int_{(m_{c}+m_{s})^{2}}^{\infty }\frac{\rho ^{%
\mathrm{QCD}}(s)}{s-p^{2}}+...,  \label{CFQCD}
\end{equation}%
where $\rho ^{\mathrm{QCD}}(s)$ is the corresponding spectral density. It
can be computed using mathematical methods described in Refs.\ \cite%
{Agaev:2016mjb,Agaev:2016dev}. Therefore, here we omit details of
calculations and provide explicit expressions for both $\rho^{\mathrm{QCD}%
}(s)$ in Appendix A.

Applying the Borel transformation on the variable $p^{2}$ to the invariant
amplitude $\Pi ^{\mathrm{QCD}}(p^{2})$, equating the obtained expression  $%
\mathcal{B}_{p^{2}}\Pi ^{\mathrm{Phys}}(p)$, and subtracting the continuum
contribution, we finally obtain the required sum rule. Thus, the mass of the
$X_{c}$ state can be evaluated from the sum rule
\begin{equation}
m_{X_{c}}^{2}=\frac{\int_{(m_{c}+m_{s})^{2}}^{s_{0}}dss \rho ^{\mathrm{QCD}%
}(s)e^{-s/M^{2}}}{\int_{(m_{c}+m_{s})^{2}}^{s_{0}}ds\rho ^{\mathrm{QCD}%
}(s)e^{-s/M^{2}}},  \label{eq:srmass}
\end{equation}%
whereas for the decay constant $f_{X_{c}}$ we employ the formula
\begin{equation}
f_{X_{c}}^{2}m_{X_{c}}^{2}e^{-m_{X_{c}}^{2}/M^{2}}=%
\int_{(m_{c}+m_{s})^{2}}^{s_{0}}ds\rho ^{\mathrm{QCD}}(s)e^{-s/M^{2}}.
\label{eq:srcoupling}
\end{equation}%

The last two expressions are the sum rules needed to evaluate the $X_{c}$
state's mass and decay constant, respectively.
For numerical computation we need values of the quark, gluon and mixed condensates.
Additionally, QCD sum rules contain $c$ and $s$ quark masses. The values of used
parameters are moved to Table \ref{tab:Param1}.

\begin{figure}[tbp]
\centerline{
\begin{picture}(210,170)(0,0)
\put(-10,0){\epsfxsize8.2cm\epsffile{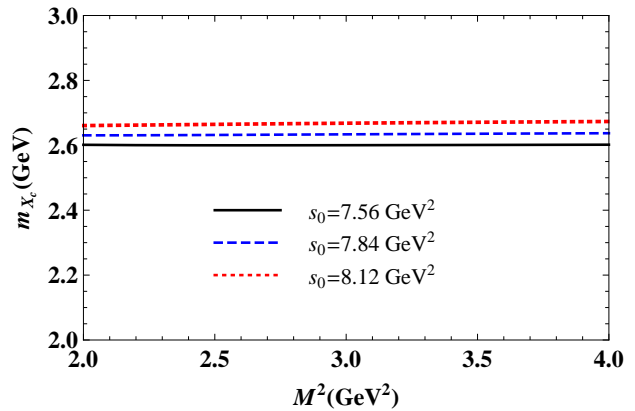}}
\end{picture}
}
\caption{The mass $m_{X_c}$ as a function of the Borel parameter $M^2$ for
different values of $s_0$. In calculations the current $J_{2}^{X_{c}}$ is
used.}
\label{fig:Xcmass}
\end{figure}

Sum rule calculations imply fixing regions for the parameters $s_{0}$ and $%
M^2$, where they can be varied. For $s_{0}$ we employ
\begin{equation}
7.56\,\,\mathrm{GeV}^2 \leq s_{0}\leq 8.12 \,\,\mathrm{GeV}^2.
\end{equation}
We find the range $2\ \mathrm{GeV}^2<M^2<4\ \mathrm{GeV}^2$ as a reliable
region for varying the Borel parameter. Here the effects of the higher
resonances and continuum states, and contributions of the higher dimensional
condensates satisfy well known requirements of QCD sum rule calculations. It
is not difficult to see that, in these intervals, the dependences of the mass
and decay constant on $M^2$ and $s_{0}$ are very weak, and we expect that the
sum rules give the firm predictions (see Figs. ~\ref{fig:Xcmass} and \ref%
{fig:Xccoup}). We estimate errors of the numerical computations by
varying the parameters $M^2$ and $s_{0}$ within the accepted ranges, as well
as taking into account uncertainties coming from other input parameters.

For the mass and decay constant of the $X_c$ state we find:
\begin{eqnarray}
&&m_{X_c}=(2590\pm 60)~\mathrm{MeV}, \notag \\
&&f_{X_c}=(0.20 \pm 0.03)\cdot 10^{-2}~\mathrm{GeV}^4,
\end{eqnarray}
when using the interpolating current $J_{1}^{X_{c}}$, and
\begin{eqnarray}
&&m_{X_c}=(2634\pm 62)~\mathrm{MeV}, \notag \\
&&f_{X_c}=(0.11 \pm 0.02)\cdot 10^{-2}~\mathrm{GeV}^4
\end{eqnarray}
in the case of $J_{2}^{X_{c}}$. As is seen, for the mass of the $X_{c}$ state
the different interpolating currents lead to predictions, which are very close to each other.
The result for the mass of the $X_c$ state obtained in Ref.\ \cite{Chen:2016mqt}
\begin{equation}
m_{X_c}=(2.55\pm 0.09)~\mathrm{GeV},
\end{equation}
within the errors is in agreement
with our predictions.

\begin{table}[tbp]
\begin{tabular}{|c|c|}
\hline\hline
Parameters & Values \\ \hline\hline
$m_{c}$ & $(1.275\pm0.025)~\mathrm{GeV}$ \\
$m_{s} $ & $(95 \pm 5)~\mathrm{MeV} $ \\
$\langle \bar{q}q \rangle $ & $(-0.24\pm 0.01)^3 $ $\mathrm{GeV}^3$ \\
$\langle \bar{s}s \rangle $ & $0.8\ \langle \bar{q}q \rangle$ \\
$\langle\frac{\alpha_sG^2}{\pi}\rangle $ & $(0.012\pm0.004)$ $~\mathrm{GeV}%
^4 $ \\
$m_{0}^2 $ & $(0.8\pm0.1)$ $\mathrm{GeV}^2$ \\
$\langle \overline{q}g\sigma Gq\rangle $ & $m_{0}^2\langle \bar{q}q \rangle$
\\ \hline\hline
\end{tabular}%
\caption{Input parameters used in calculations.}
\label{tab:Param1}
\end{table}
\begin{figure}[tbp]
\centerline{
\begin{picture}(200,170)(0,0)
\put(-10,20){\epsfxsize8.2cm\epsffile{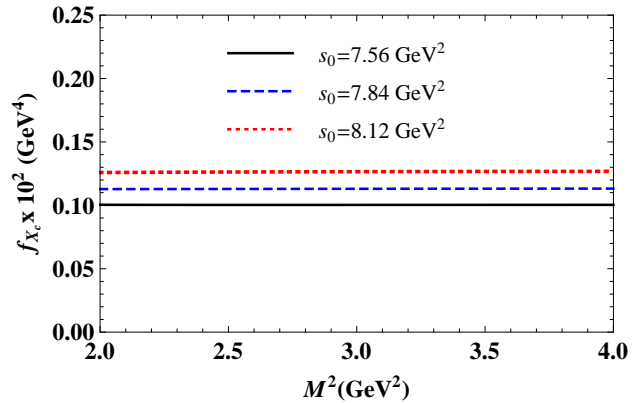}}
\end{picture}}
\caption{The decay constant $f_{X_{c}}$ vs Borel parameter $M^2$ for $%
J_{2}^{X_{c}}$. The values of the parameter $s_0$ are shown in the figure.}
\label{fig:Xccoup}
\end{figure}

\section {THE strong decays of the $X_{c}$ state}

\label{sec:Vertex}
Predictions for the mass of the $X_c$ state obtained in the previous section allow us to
continue our exploration by considering its possible decay channels and to calculate their
decay widths. From the quark content and assigned quantum numbers,  it is easy
to conclude that the $X_c$ state can decay into 
$D_{s}^{-}(s\bar {c}) + \pi^{+}(u\bar {d})$ or $D^{0}(u\bar {c}) + K^{0}(s\bar{d})$. In other words,
$X_{c}\rightarrow D_{s}^{-}\pi ^{+}$ and $X_{c}\rightarrow D^{0}K^{0}$ transitions are kinematically allowed decay channels of the $X_c$ state. Our aim in this section is to find   widths of these decays. To this end, we calculate the
strong couplings  $g_{X_{c}D_{s}\pi }$ and $g_{X_{c}DK}$ using the method of
QCD sum rule on the light-cone in conjunction with the soft-meson
approximation \cite{Agaev:2016dev}.

We start our analysis from the decay $X_c \to D_{s}^{-} \pi^{+} $ . In order to calculate the required  strong coupling $g_{X_cD_s \pi }$ we consider the correlation function
\begin{equation}
\Pi(p,q)=i\int d^{4}xe^{ipx}\langle \pi (q)|\mathcal{T}%
\{J^{D_{s}}(x)J^{X_{c}\dag }(0)\}|0\rangle.  \label{eq:CorrF3}
\end{equation}
Here the interpolating current $J^{X_{c}}(x)$ is given by Eq.\ (\ref{eq:CDi2}%
), whereas for $D_{s}^{-}$ we use
\begin{equation}
J^{D_s}(x)=\overline{c}^{l}(x)i\gamma _{5 }s^{l}(x).  \label{eq:Bcur}
\end{equation}

It is not difficult to find $\Pi (p,q)$ in terms of the physical degrees of
freedom:
\begin{eqnarray}
\Pi ^{\mathrm{Phys}}(p,q) &=&\frac{\langle 0|J^{D_{s}}|D_{s}\left( p\right)
\rangle }{p^{2}-m_{D_{s}}^{2}}\langle D_{s}\left( p\right) \pi
(q)|X_{c}(p^{\prime })\rangle  \notag \\
&&\times \frac{\langle X_{c}(p^{\prime })|J^{X_{c}\dagger }|0\rangle }{%
p^{\prime 2}-m_{X_{c}}^{2}}+\ldots ,  \label{eq:CorrF4}
\end{eqnarray}%
where by dots we denote contributions of the higher resonances and continuum
states. Here $p$, $q$ and $p^{\prime }=p+q$, are the momenta of $D_{s}$, $\pi
$, and $X_{c}$ states, respectively. In order to finish computation of
the correlation function we introduce the matrix elements
\begin{eqnarray}
&&\langle 0|J^{D_{s}}|D_{s}\left( p\right) \rangle =\frac{%
f_{D_{s}}m_{D_{s}}^{2}}{m_{c}+m_{s}},  \notag \\
&&\langle X_{c}(p^{\prime })|J^{X_{c}\dagger }|0\rangle =f_{X_{c}}m_{X_{c}},
\notag \\
&&\langle D_{s}\left( p\right) \pi (q)|X_{c}(p^{\prime })\rangle
=g_{X_{c}D_{s}\pi }p\cdot p^{\prime },  \label{eq:Mel}
\end{eqnarray}
where $f_{X_{c}}$ and $m_{X_{c}}$ are the decay constant and mass of the $X_{c}$
state, whereas $f_{D_{s}}$ and $m_{D_{s}}$ are the same parameters of the $D_{s}$
meson.

We calculate $\Pi ^{\mathrm{Phys}}(p,q)$ in the soft-meson limit $q=0$, and
after some manipulations described in Refs.\ \cite%
{Agaev:2016ijz,Agaev:2016dev}, for the Borel transformation of the correlation
function find
\begin{eqnarray}
&&\Pi ^{\mathrm{Phys}}(M^{2})=\frac{%
f_{D_{s}}f_{X_{c}}m_{X_{c}}m_{D_{s}}^{2}g_{X_{c}D_{s}\pi }}{(m_{c}+m_{s})}%
m^{2}  \notag \\
&&\times \frac{1}{M^{2}}e^{-m^{2}/M^{2}},
\end{eqnarray}%
where $m^2=(m_{X_c}^{2}+m_{D_{s}}^{2})/2$.

To proceed, we have to calculate $\Pi ^{\mathrm{QCD}}(p,q)$  in terms of the quark-gluon
degrees of freedom and find  QCD side of the sum rule. Contractions
 of  $s$ and $c$-quark fields in Eq.\ (\ref{eq:CorrF3}) yield
\begin{eqnarray}
&&\Pi ^{\mathrm{QCD}}(p,q)=\int d^{4}xe^{ipx}\varepsilon ^{ijk}\varepsilon
^{imn}\left[ \gamma_{5}\widetilde{S}_{s}^{lj}(x){}\gamma _{5}\right.
\notag \\
&&\left. \times \widetilde{S}_{b}^{ml}(-x){}\gamma _{5}\right] _{\alpha
\beta }\langle \pi (q)|\overline{u}_{\alpha }^{k}(0)d_{\beta
}^{n}(0)|0\rangle ,  \label{eq:CorrF6}
\end{eqnarray}%
where $\alpha $ and $\beta $ are the spinor indexes.

Skipping technical details, which can be found in
Refs.\ \cite{Agaev:2016ijz,Agaev:2016dev},
we provide final expression for the spectral density, which is given as a sum of
the perturbative and nonperturbative components
\begin{equation}
\rho_{\mathrm{coup.}} ^{\mathrm{QCD}}(s)=\rho ^{\mathrm{pert.}}(s)+\rho ^{\mathrm{n.-pert.}%
}(s).  \label{eq:SD}
\end{equation}%
where
\begin{equation}
\rho ^{\mathrm{pert.}}(s)=\frac{f_{\pi }\mu _{\pi }}{16\pi ^{2}s}\sqrt{%
s(s-4m_{c}^{2})}\left( s+2m_{c}m_{s}-2m_{c}^{2}\right) ,  \label{eq:SD1}
\end{equation}%
and
\begin{eqnarray}
&&\rho ^{\mathrm{n.-pert.}}(s)=\frac{f_{\pi }\mu _{\pi }}{72}\left\{
6\langle \overline{s}s\rangle \left[ -2m_{c}\delta (s-m_{c}^{2})\right.
\right.   \notag \\
&&\left. +sm_{s}\delta ^{^{(1)}}(s-m_{c}^{2})\right] +\langle \overline{s}%
g\sigma Gs\rangle \left[ 6(m_{c}-m_{s})\delta ^{^{(1)}}(s-m_{c}^{2})\right.
\notag \\
&&\left. \left. -3s(m_{c}-2m_{s})\delta ^{(2)}(s-m_{c}^{2})-s^{2}m_{s}\delta
^{(3)}(s-m_{c}^{2})\right] \right\} .  \label{eq:SD2}
\end{eqnarray}%
In Eq.\ (\ref{eq:SD2}) $\delta ^{(n)}(s-m_{c}^{2})=(d/ds)^{n}\delta
(s-m_{c}^{2})$ that appear when extracting the imaginary part of the pole
terms.

As is seen, in the soft limit the spectral density depends only the
parameters $f_{\pi}$ and $\mu_{\pi}$ through the pion's local matrix element
\begin{equation}
\langle 0|\overline{d}(0)i\gamma _{5}u(0)|\pi (q)\rangle =f_{\pi }\mu _{\pi
},  \label{eq:MatE2}
\end{equation}%
where
\begin{equation}
\mu _{\pi }=\frac{m_{\pi }^{2}}{m_{u}+m_{d}}=-\frac{2\langle \overline{q}%
q\rangle }{f_{\pi }^{2}}.  \label{eq:PionEl}
\end{equation}%

Continuum subtraction performed in the standard way leads to the final sum
rule for evaluating of the strong coupling
\begin{eqnarray}
&&g_{X_{c}D_{s}\pi }=\frac{(m_{c}+m_{s})}{%
f_{D_{s}}f_{X_{c}}m_{X_{c}}m_{D_{s}}^{2}m^{2}}\left( 1-M^{2}\frac{d}{dM^{2}}%
\right) M^{2}  \notag \\
&&\times \int_{(m_{c}+m_{s})^{2}}^{s_{0}}dse^{(m^{2}-s)/M^{2}}\rho ^{\mathrm{%
QCD}}(s).  \label{eq:SRules}
\end{eqnarray}

The width of the decay $X_{c}\rightarrow D_{s}^{-}\pi ^{+}$ can be found
applying the standard methods and is given in Ref.\ \cite{Agaev:2016ijz}:
\begin{eqnarray}
&&\Gamma \left( X_{c}\rightarrow D_{s}^{-}\pi ^{+}\right) =\frac{%
g_{X_{c}D_{s}\pi }^{2}m_{D_{s}}^{2}}{24\pi }\lambda \left( m_{X_{c}},\
m_{D_{s}},m_{\pi }\right)   \notag \\
&&\times \left[ 1+\frac{\lambda ^{2}\left( m_{X_{c}},\ m_{D_{s}},m_{\pi
}\right) }{m_{D_{s}}^{2}}\right] ,  \label{eq:DW}
\end{eqnarray}%
where
\begin{equation*}
\lambda (a,\ b,\ c)=\frac{\sqrt{a^{4}+b^{4}+c^{4}-2\left(
a^{2}b^{2}+a^{2}c^{2}+b^{2}c^{2}\right) }}{2a}.
\end{equation*}%
Equations\ (\ref{eq:SRules}) and (\ref{eq:DW}) are final expressions that
will be used for numerical analysis of the decay channel $X_{c}\rightarrow
D_{s}^{-}\pi ^{+}$.

The investigation of the transition $X_{c}\rightarrow
D^{0}K^{0}$ can be carried out in the same manner as for the decay $X_{c}\rightarrow
D_{s}^{-}\pi ^{+}$. One needs only to replace the parameters of the particles in accordance with
the prescription $\pi \to K$, and $D_s \to D$. Nevertheless, below we write
down some key expressions.

Thus, the analysis of the vertex $X_{c}D K$, which is necessary to derive the sum rule
for the coupling $g_{X_{c}D K }$, is founded on the correlation function
\begin{equation}
\Pi_{K}(p,q)=i\int d^{4}xe^{ipx}\langle K (q)|\mathcal{T}\{J^{D}(x)J^{X_{c}%
\dag }(0)\}|0\rangle,  \label{eq:CorrF3A}
\end{equation}
where for $D^{0}$ meson we employ the interpolating current
\begin{equation}
J^{D}(x)=\overline{c}^{l}(x)i\gamma _{5 }u^{l}(x).  \label{eq:BcurA}
\end{equation}
In the soft-meson limit $q=0$, the Borel transformation of the correlation function
$\Pi _{K}^{\mathrm{Phys}}(p,q)$ is given by
\begin{eqnarray}
&&\Pi _{K}^{\mathrm{Phys}}(M^{2})=\frac{%
f_{D}f_{X_{c}}m_{X_{c}}m_{D}^{2}g_{X_{c}DK}}{(m_{c}+m_{u})}m^{2}  \notag \\
&&\times \frac{1}{M^{2}}e^{-m^{2}/M^{2}}.
\end{eqnarray}%
In the formula above $m^{2}=(m_{X_{c}}^{2}+m_{D}^{2})/2$, and $m_{D}$ and $f_{D}$
are  the mass and decay constant of D meson, respectively.

In terms of the quark-gluon degrees of freedom the same function
is determined by means of the formula
\begin{eqnarray}
&&\Pi _{K}^{\mathrm{QCD}}(p,q)=\int d^{4}xe^{ipx}\varepsilon
^{ijk}\varepsilon ^{imn}\left[ \gamma _{5}\widetilde{S}_{u}^{lj}(x){}\gamma
_{5}\right.   \notag \\
&&\left. \times \widetilde{S}_{c}^{ml}(-x){}\gamma _{5}\right] _{\alpha
\beta }\langle K(q)|\overline{s}_{\alpha }^{k}(0)d_{\beta }^{n}(0)|0\rangle .
\end{eqnarray}%
Its imaginary part gives us the spectral density $\rho_{\mathrm{coup.}} ^{\mathrm{QCD}}(s)$,
which now depends on the K meson local matrix element
\begin{equation}
\langle 0|\overline{d}(0)i\gamma _{5}s(0)|K(q)\rangle
=\frac{f_{K}m_{K}^{2}}{m_{s}+m_{d}},
\end{equation}%
with $m_{K}$ and $f_{K}$ being the  mass and decay constant of the K meson.
The remaining analysis is the same as for the  $X_{c}\rightarrow D_{s}^{-}\pi ^{+}$ decay:
after evident changes in the relevant final expressions, they can be utilized
for studying of the $X_{c}\rightarrow DK$ transition, as well.

The QCD sum rules derived above contain, as input parameters,  the masses and decay constants
of the $D_{s}$, $D$, $\pi$ and $K$ mesons. They are collected in  Table \ref{tab:Param}.
It is worth noting that for the decay constants $f_{D}$ and $f_{D_s}$ we use the lattice result
from Ref.\ \cite{Bazavov1:2011aa}.

The results of the numerical calculations of the strong couplings and decay widths are shown in Table \ref{tab:Results}.
We find that the transition $X_{c}\rightarrow DK$ may be viewed as the dominant decay channel of the
$X_c$ state. The total width of this particle computed by taking into account the explored decay channels
equals to
\begin{equation}
\Gamma_{X_c}^{1}\simeq (63.4  \pm 14.2)\, \mathrm{MeV},\,
\end{equation}
and
\begin{equation}
\Gamma_{X_c}^{2}\simeq (53.7  \pm 11.6 )\, \mathrm{MeV},
\end{equation}
for the first and second interpolating currents, respectively.
It is seen that results obtained for the total width of the $X_c$  state
using  various interpolating currents, within errors, are compatible with each other,
nevertheless the difference between the central values are sizeable.
The experimental exploration of the $X_c$ state, and its observation may extend our
knowledge about the nature and internal structure of the new exotic states.
\begin{table}[tbp]
\begin{tabular}{|c|c|}
\hline\hline
Parameters & Values \\ \hline\hline
$m_{D_s}$ & $(1968.30\pm 0.10)~\mathrm{MeV}$ \\
$f_{D_s}$ & $(260.1\pm 10.8)~\mathrm{MeV}$ \\
$m_{D}$ & $(1864.84\pm 0.05)~\mathrm{MeV}$ \\
$f_{D}$ & $(218.9\pm 11.3)~\mathrm{MeV}$ \\
$m_{K}$ & $497.61 ~\mathrm{MeV}$ \\
$f_{K}$ & $156~\mathrm{MeV}$ \\
$m_{\pi}$ & $139.57 ~\mathrm{MeV}$ \\
$f_{\pi}$ & $131~\mathrm{MeV}$ \\ \hline\hline
\end{tabular}%
\caption{Input parameters used in the coupling calculations.}
\label{tab:Param}
\end{table}
\begin{table}[tbp]
\begin{tabular}{|c|c|c|}
\hline\hline
& $J_1^{X_c}$ & $J_2^{X_c}$ \\ \hline\hline
$g_{X_{c}D_{s}\pi}$ & $(0.51\pm 0.10) $ $\mathrm{GeV}^{-1}$ & $(0.51\pm
0.11) $ $\mathrm{GeV}^{-1}$ \\
$\Gamma_{D_{s} \pi}$ & $(8.0 \pm 2.0)$ $\mathrm{MeV}$ & $(8.2 \pm 2.1)$ $\mathrm{%
MeV}$ \\
$g_{X_{c}D K}$ & $(1.57 \pm 0.34) $ $\mathrm{GeV}^{-1}$ & $(1.36 \pm 0.32) $ $%
\mathrm{GeV}^{-1}$ \\
$\Gamma_{D K}$ & $(55.4 \pm 14.0)$ $\mathrm{MeV}$ & $(45.5 \pm 11.4)$ $\mathrm{MeV}$
\\ \hline\hline
\end{tabular}%
\caption{The sum rule predictions for the strong couplings and corresponding
decay widths.}
\label{tab:Results}
\end{table}

\section*{ACKNOWLEDGEMENTS}

The work of S.~S.~A. was supported by the TUBITAK grant 2221-"Fellowship
Program For Visiting Scientists and Scientists on Sabbatical Leave". This
work was also supported in part by TUBITAK under the grant no: 115F183.

\appendix*

\section{A}

\renewcommand{\theequation}{\Alph{section}.\arabic{equation}}

\label{sec:App} In this appendix we have collected the results of our
calculations of the spectral density
\begin{equation}
\rho ^{\mathrm{QCD}}(s)=\rho ^{\mathrm{pert}}(s)+\sum_{k=3}^{k=8}\rho
_{k}(s),  \label{eq:A1}
\end{equation}%
used for evaluation of the $X_{c}$ meson mass $m_{X_{c}}$ and its decay
constant $f_{X_{c}}$ from the QCD sum rule. In Eq.\ (\ref{eq:A1}) by $\rho
_{k}(s)$ we denote the nonperturbative contributions to $\rho ^{\mathrm{QCD}%
}(s)$. In calculations we have neglected the masses of the $u$ and $d$
quarks and taken into account terms $\sim m_s$. The explicit expressions for
$\rho ^{\mathrm{pert}}(s)$ and $\rho _{k}(s)$ in the case of the current $%
J_{2}^{X_{c}}(x)$ are presented below as integrals over the Feynman
parameter $z$. Note that in $\rho_8(s)$, we keep only the term containing the
gluonic contribution.
\begin{widetext}
\begin{eqnarray}
&&\rho ^{\mathrm{pert}}(s) =\frac{1}{6144\pi ^{6}}\int\limits_{0}^{a}\frac{%
dzz^{4}}{(z-1)^{3}}\left[ m_{c}^{2}+s(z-1)\right] ^{3}\left[
m_{c}^{2}+3s(z-1)\right] , \notag \\
&&\rho _{\mathrm{3}}(s) =\frac{1}{64\pi ^{4}}\int\limits_{0}^{a}\frac{dzz^{2}%
}{(z-1)^{2}}\left[ m_{c}^{2}+s(z-1)\right] \left\{ \langle \overline{d}%
d\rangle m_{c}\left[ m_{c}^{2}+s(z-1)\right] +m_{s}(\langle \overline{s}%
s\rangle -2\langle \overline{u}u\rangle )\left[ m_{c}^{2}+2s(z-1)\right]
(z-1)\right\} , \notag \\
&&\rho _{\mathrm{4}}(s) =\frac{1}{9216\pi ^{4}}\langle \alpha _{s}\frac{G^{2}%
}{\pi }\rangle \int\limits_{0}^{a}\frac{dzz^{2}}{(z-1)^{3}}\left\{ 2m_{c}^{4}%
\left[ z(7z-15)+9\right] +3m_{c}^{2}s(z-1)\left[ z(13z-30)+18\right]
+12s^{2}(z-1)^{3}(2z-3)\right\} , \notag \\
&&\rho _{\mathrm{5}}(s) =\frac{m_{0}^{2}}{192\pi ^{4}}\int\limits_{0}^{a}%
\frac{dzz}{(1-z)}\left\{ 3m_{c}\langle \overline{d}d\rangle \left[
m_{c}^{2}+s(z-1)\right] +m_{s}(z-1)(\langle \overline{s}s\rangle -3\langle
\overline{u}u\rangle )\left[ 2m_{c}^{2}+3s(z-1)\right] \right\} , \notag \\
&&\rho _{\mathrm{6}}(s) =\frac{g^{2}}{1296\pi ^{4}}\int\limits_{0}^{a}dzz(%
\langle \overline{u}u\rangle ^{2}+\langle \overline{d}d\rangle ^{2}+\langle
\overline{s}s\rangle ^{2})\left[ 2m_{c}^{2}+3s(z-1)\right] , \notag \\
&&\rho _{\mathrm{7}}(s) =\frac{1}{576\pi ^{2}}\langle \alpha _{s}\frac{G^{2}%
}{\pi }\rangle \int\limits_{0}^{a}dz\left\{ 4m_{c}\langle \overline{d}%
d\rangle +m_{s}\left[ \langle \overline{u}u\rangle(4z+2) -3z\langle
\overline{s}s\rangle\right] \right\} , \notag \\
&&\rho _{\mathrm{8}}(s)=-\frac{11}{36864\pi ^{2}}\langle \alpha _{s}\frac{%
G^{2}}{\pi }\rangle ^{2}\int\limits_{0}^{a}dzz,
\label{eq:ro2}
\end{eqnarray}
\end{widetext}
where $a=(s-m_{c}^{2})/s$.

\end{document}